\documentstyle[12pt]{article}
\font\tenBbb=msbm10 \newfam\Bbbfam \textfont\Bbbfam=\tenBbb
\def\Bbb#1{{\fam\Bbbfam #1}}
\makeatletter\@addtoreset{equation}{section}\makeatother

\begin{document}

\begin{center}
{\Large THERMODYNAMIC LIMIT IN CHERN-SIMONS }
\\ 
{\Large SYSTEM OF PARTICLES WITH MB STATISTICS.}
\vskip 7 pt
                     W.I.S k r y p n i k
\vskip 7 pt
 Institute of mathematics, Tereshchenkivska str.3, Kyiv-4, Ukraine, 252601

\end{center}
\vskip 7 pt
 A b s t r a c t
\vskip 3 pt
The reduced density matrices (RDMs) are calculated in the thermodynamic limit
for the Chern-Simons non-relativistic particle system and Maxwell-Boltzmann
(MB) statistics. It is established that they are zero outside of a diagonal 
and well-behaved after a renormalization, depending on an arbitrary real 
number, if the condition of neutrality holds.

\vskip 20 pt
\section{Introduction.}

The Chern-Simons 2-d quantum system of n nonrelativistic spinless
identical particles of unit mass is described by the Hamiltonian $\dot{H}_n$,
defined on $C^{\infty}(\Bbb R^{2n}\backslash \cup_{j,k} (x_j=x_k))$

\begin{equation}
 \dot H_{n}= \frac{1}{2}\sum_{j=1}^{n}||p_{j} - a_{j}(X_{n}||^{2},
\hskip 20 pt
a_{j}^{\nu}(X_{n})=\epsilon^{\nu \mu}\partial_{\mu,j}U_C(X_n),\hskip 10 pt
\nu,\mu=1,2,
\end{equation}
\vskip 5 pt
$$
U_C(X_n)=g\sum\limits_{1\leq k<j\leq n}e_je_kln|x_j-x_k|,\hskip 10 pt
X_{n}=(x_{1},...x_{n})\in \Bbb R^{2n},\hskip 10 pt
p^{\mu}_{j}=i^{-1}\partial_{\mu,j}=i^{-1}\frac{\partial}{\partial x^{\mu}_j},
$$
where $||v||^2=(v^1)^2+(v^2)^2$,  $\epsilon$ is the skew symmetric tensor 
and the repeating index implies a summation, real number $e_j$(a charge) 
takes values in a finite set $E_{\{c\}}$ from $\Bbb R$.

Differentiating the equality $f(f^{-1}(x))=x$ we derive the formula
$$
\frac{df^{-1}(x)}{dx}=(\frac{df(y)}{dy})^{-1}, \hskip 20 pt y=f^{-1}(x).
$$
From this equality for $f(x)=\tan x$ and the equality \hskip 10 pt
$\frac{d}{dx}\tan x=1+\tan^2(x)$\hskip 10 pt the following relation is derived

$$
\frac{\partial}{\partial x^{\nu}} \arctan
\frac{x^2}{x^1}=\epsilon^{\nu\mu}x^{\mu}|x|^{-2}=
\epsilon^{\nu\mu}\frac{\partial}{\partial x^{\mu}}ln |x|. 
$$

This means that CS system is almost(quasi-)integrable, that is

$$
\dot{H}_n=e^{i\hat{U}}\dot{H}^0_ne^{-i\hat{U}},
$$
 where $-2\dot{H}^0_n$ is the 2n-dimensional Laplacian $-2H_n^0$ restricted to
$C^{\infty}(\Bbb R^{2n}\backslash \cup_{j,k} (x_j=x_k))$ and $\hat{U}$ is the 
operator of multiplication by $U(X_n)$,

$$
U(X_n)=g\sum\limits_{1\leq k<j\leq n}e_je_k\phi(x_j-x_k), \hskip 20 pt
\phi(x_j-x_k)=\arctan \frac{x^2_j-x^2_k}{x^1_j-x^1_k}
$$
As a result, there exists the simplest sefadjoint extension $H_n$ of 
$\dot{H}_n$

\begin{equation}
H_n=e^{i\hat{U}}H^0_ne^{-i\hat{U}}.
\end{equation}

with the domain $D(H_n)=e^{i\hat{U}}D(H_n^0)$. Another selfadjoint extension 
of $\dot{H_n}$ is given by (1.2) in which, instead of $H_n^0$, another 
selfadjoint extension of $\dot{H}^0_n$ is considered.

The CS system of particles with different statistics has been studied by many
authors [1-4] since it can be derived (formally) in
3-d topological electrodynamics (its Lagrangian contains CS term) in the limit 
of the vanishing Maxwell term ( the same is true for its relativistic version).
There is a hope it can give a new mechanism of superconductivity, 
superfluidity and $P,T$ violation (general outlook of modern statistical 
mechanics allows to expect breakdown only of discrete (hidden) symmetries 
for quantum systems at nonzero temperature in 3 space-time dimensions).

In this paper the CS system is considered with the Hamiltonian (1.2) and the 
Maxwell-Boltzmann (MB) statistics. Its thermodynamics coincides with 
thermodynamics of the free particle system if Dirichlet boundary condition is 
considered(we do this in this paper). Our calculations show that the
Gibbs (grand-canonical)reduced density matrices (RDMs) 
$\rho^{B_L}(X_m|Y_m)$ in the sphere $B_L$ of radius L, centered at the 
origin, tend to zero if $x_j\not= y_j$. They are expressed as the product of a 
 function having a non-zero limit and  

$$
\exp\{-\frac{N_{r,L}}{4}g^2(\sum\limits_{e}z_ee^2)
\sum\limits_{j=1}^{m}e^2_j|x_j-y_j|^2\}, \hskip 20 pt
N_{r,L}=\int\limits_{r<|x|\leq L}
P_{0(B_L)}^{\beta}(x|x)|x|^{-2}dx.
$$

where $z_e$ is the activity of the particles with the charge $e$ and $N_{r,L}$ 
diverges as $2ln L$, $P_{0(B_L)}^{\beta}(x|y)$ is the kernel of the 
semigroup whose infinitesimal generator coincides with the two-dimensional 
Laplacian with the Dirichlet boundary condition on the boundary of the ball 
$B_L$. We propose to renormalize the RDMs by dividing them by 
this exponent. We show then the important feature of this renormalization: 
if the condition of the neutrality holds then the resulting RDMs in the 
thermodynamic limit $\rho(X_m|Y_m)$ satisfy the compatibility 
condition( see the corollary), that is, they define a state of the infinite 
particle system.

The unusual property of the RDMs to be zero almost everywhere was found by us 
in 1-d (integrable) systems of particles with the Hamiltinian (1.1) when $a_j$ 
is expressed through a pair (scalar magnetic) long-range potential [5-6]. We
expect that the similar property is true for bosonic and fermiomic CS systems
for small values of activities of particles(see also [7]).

The results of our calculations are formulated in the theorem  and corollary 
in the next paragragh.
\vskip 20 pt

\section{Main result.}

Let $\Lambda\in \Bbb R^2$ be a compact set  and assume the Dirichlet boundary 
conditions on the boundary $\partial \Lambda$. For the inverse 
tempetature $\beta$, and the activity $z_{e_s}$ of the particles with the 
charge $e_s$, the Gibbs (grand-canonical, equilibrium) RDMs are given by 
$$
\rho^{\Lambda}(X_m|Y_m)=
$$
$$
=Z_{(e)_m}\Xi^{-1}_{\Lambda}
\sum\limits_{n\geq0} \prod\limits_{s=1}^{n}\sum\limits_{e_s'}
\frac{z_{e_s'}^{n_s}}
{n_s!} \int\limits_{\Lambda^n} dX'_n \exp\{i[
U(X_m,X_n')-U(Y_m,X_n')
]\}P^{\beta}_{0(\Lambda)}(X_m,X_n'|Y_m,X_n'),
$$
where $\Xi_{\Lambda}$ is the grand partition function (it coincides with the 
numerator in the r.h.s. of this equality for the case "m=0", i.e. when there 
are no $X_m$ and $Y_m$), 
$P^{\beta}_{0(\Lambda)}(X_m|Y_m)$ is the kernel of the semigroup, 
whose infinitesimal generator concides with  $H^0_{n,\Lambda}$ 
($-2H^0_{n,\Lambda}$ is the Laplacian in $\Lambda^n$ with the 
Dirichlet boundary condition on the boundary $\partial \Lambda^n$),
$Z_{(e)_m}=\prod\limits_{s=1}^{m}z_{e_s}$, and the summation in $e_s$ is
performed over $E_{\{c\}}$.
\begin{equation}
P^{\beta}_{0(\Lambda)}(X_n|Y_n)=\prod\limits_{j=1}^{n}P^{\beta}_
{0(\Lambda)}(x_j|y_j),
\hskip 7 pt X_m=(X_m^1,X_m^2), Y_m=(Y_m^1,Y_m^2) \in \Bbb R^{2m},
\end{equation}
$P^{\beta}_{0(\Lambda)}(x|y)$ 
is the transition probability  of the 2-dimensional free diffusion proccess 
with the Dirichlet boundary condition on $\partial\Lambda$.
$$
P^{\beta}_{0(\Lambda)}(x|y)=\int P_{x,y}^{\beta}(d\omega)\chi_{\Lambda}
(\omega),
$$
$P_{x,y}(d\omega)$ is the conditional Wiener measure and $\chi_{\Lambda}
(\omega)$ is the characteristic function of the paths that are inside 
$\Lambda$.

From the equality
$$
U(X_m,X_n')
=U(X_m)+\sum\limits_{j=1}^{m}\sum\limits_{k=1}^{n}\phi(x_j-x_k')e_je_k'
+U(X_n')
$$
we obtain 

$$
\rho^{\Lambda}(X_m|Y_m)=
Z_{(e)_m}\Xi_{\Lambda}^{-1}\exp\{i[U(X_m)-U(Y_m)]\}
P^{\beta}_{0(\Lambda)}(X_m|Y_m)\sum\limits_{n\geq0}
\sum\limits_{e_s'}\frac{z_{e_s'}^{n_s}}
{n_s!} \times
$$
$$
\times \int\limits_{\Lambda^n} dX'_n \prod\limits_{k=1}^{n}
\exp\{i\sum\limits_{j=1}^{m}e_je_k(\phi(x_j-x_k')-
\phi(y_j-x_k'))\}P^{\beta}_{0(\Lambda)}(x_k'|x_k'),
$$

As a result

\begin{equation}
\rho^{\Lambda}(X_m|Y_m)=Z_{(e)_m}\exp\{i[U(X_m)-U(Y_m)]+
G_{\Lambda}(X_m|Y_m)\}P^{\beta}_{0(\Lambda)}(X_m|Y_m),
\end{equation}

\begin{equation}
G_{\Lambda}(X_m|Y_m)=\sum\limits_{e}z_e\int\limits_{\Lambda}
P_{0(\Lambda)}^{\beta}(x|x)
[\exp\{i\sum\limits_{j=1}^{m}ee_j(\phi(x_j-x)-
\phi(y_j-x))\}-1]dx.
\end{equation}

Here we used the equality
$\Xi_{\Lambda}=
\exp\{\sum\limits_{e}z_e\int\limits_{\Lambda}P_{0(\Lambda)}^{\beta}(x|x)dx\}$.

With the help of the equality
$$
exp\{i\arctan \frac{x^2}{x^1}\}=\frac{x}{|x|}=(\frac{x}{x^*})^{\frac{1}{2}},
\hskip 20 pt x=x^1+ix^2,
$$
we derive
$$
\exp\{i\sum\limits_{j=1}^{m}ee_j(\phi(x_j-x)-\phi(y_j-x))\}=\prod\limits_{j=1}
^{m}\left(\frac{(x-x_j)(x^*-y_j^*)}{(x^*-x_j^*)(x-y_j)}\right)^{\frac{1}{2}gee_j}=G_x(X_m|Y_m).
$$

We have to use the Taylor expansions for $|x|<1$
$$
(1-x)^g=1-gx+\frac{g(g-1)}{2}x^2+\sum\limits_{n\geq 3}C^g_nx^n=
\sum\limits_{n\geq 0}C^g_nx^n,
$$

As a result for $g'=\frac{1}{2}gee_j, g'\not\in \Bbb Z,
|\frac{x_j}{x}|<1, |\frac{y_j}{x}|<1$

$$
\left(\frac{x-x_j}{x^*-x^*_j}\right)^{g'}=\left(\frac{x}{x^*}\right)
^{g'}\left(\frac{1-\frac{x_j}{x}}{1-\frac{x^*_j}{x^*}}\right)^{g'}=
\left(\frac{x}{x^*}\right)
^{g'}\{1+g'(-\frac{x_j}{x}+\frac{x_j^*}{x^*})+
$$

$$
+\frac{g'(g'-1)}{2}
(\frac{x_j^2}{x^2}+\frac{x_j^{*2}}{x^{*2}})-g'^2|\frac{x_j}{x}|^2+
\sum\limits_{n_1^+
n_1^-\geq 3}C^{-g'}_{n_1^+}C^{g'}_{n_1^-}(\frac{x^*_j}{x^*})^{n_1^+}
(\frac{x_j}{x})^{n_1^-}\},
$$

Applying this formula we deduce
$$
\left(\frac{(x-x_j)(x^*-y^*_j)}{(x^*-x^*_j)(x-y_j)}\right)^{g'}=
$$

$$
=1+G_x'(x_j|y_j)+\sum\limits_{n_1^++...+n_2^-\geq 3}
C^{-g'}_{n_1^+}C^{g'}_{n_1^-}C^{-g'}_{n_2^+}C^{g'}_{n_2^-}
(\frac{x^*_j}{x^*})^{n_1^+}(\frac{x_j}{x})^{n_1^-}(\frac{y^*_j}{x^*})^{n_2^-}
(\frac{y_j}{x})^{n_2^+}.
$$
where

$$
G'_x(x_j|y_j)=g'(-\frac{x_j-y_j}{x}+\frac{x_j^*-y_j^*}{x^*})-
g'^2(\frac{x_j}{x}-\frac{x_j^*}{x^*})(\frac{y_j}{x}-\frac{y_j^*}{x^*})+
$$

$$
+\frac{g'(g'-1)}{2}
(\frac{x_j^2+y_j^2}{x^2}+\frac{x_j^{*2}+y_j^{*2}}{x^{*2}})
-g'^2[|\frac{x_j}{x}|^2+|\frac{y_j}{x}|^2].
$$

As a result

\begin{equation}
G_x(X_m|Y_m)=1+G^0_{x,e}(X_m|Y_m)+\sum\limits_{j=1}^{m}G_x'(x_j|y_j),
\end{equation}
where

$$
G^0_{x,e}(X_m|Y_m)=\sum\limits_{n_{1,1}^++...+n_{2,m}^-\geq
3}\prod\limits_{j=1}^{m}
C^{-g'}_{n_{1,j}^+}C^{g'}_{n_{1,j}^-}C^{-g'}_{n_{2,j}^+}C^{g'}_{n_{2,j}^-}
(\frac{x^*_j}{x^*})^{n_{1,j}^+}(\frac{x_j}{x})^{n_{1,j}^-}
(\frac{y^*_j}{x^*})^{n_{2,j}^-}(\frac{y_j}{x})^{n_{2,j}^+}.
$$

It can be checked that
$$
-g'^2(\frac{x_j}{x}-\frac{x_j^*}{x^*})(\frac{y_j}{x}-\frac{y_j^*}{x^*})
+\frac{g'^2}{2}
(\frac{x_j^2+y_j^2}{x^2}+\frac{x_j^{*2}+y_j^{*2}}{x^{*2}})
-g'^2[|\frac{x_j}{x}|^2+|\frac{y_j}{x}|^2]=
$$

\begin{equation}
=-g'^2|\frac{x_j-y_j}{x}|^2+\frac{g'^2}{2}
(\frac{(x_j-y_j)^2}{x^2}+\frac{(x_j^*-y_j^*)^2}{x^{*2}}).
\end{equation}

This yields

$$
G_x'(x_j|y_j)=-g'^2|\frac{x_j-y_j}{x}|^2 +G^-_x(x_j|y_j),\hskip 20 pt
G^-_x(x_j|y_j)=
$$
\begin{equation}
=g'(-\frac{x_j-y_j}{x}+\frac{x_j^*-y_j^*}{x^*})+\frac{g'^2}{2}
(\frac{(x_j-y_j)^2}{x^2}+\frac{(x_j^*-y_j^*)^2}{x^{*2}})
-\frac{g'}{2}
(\frac{x_j^2+y_j^2}{x^2}+\frac{x_j^{*2}+y_j^{*2}}{x^{*2}}).
\end{equation}

Let $l^+_m=max(|x_j|, |y_j|, j=1,..., m)$. Let $\Lambda=B_L$ then

$$
G_{\Lambda}(X_m|Y_m)=\sum\limits_{e}z_e\{\int\limits_{|x|\leq 2l^+_m}
P_{0(\Lambda)}^{\beta}(x|x)[\prod\limits_{j=1}^{m}
\left(\frac{x^*-x_j^*}{x-x_j}\frac{x-y_j}{x^*-y^*_j}\right)^{\frac{1}{2}gee_j}-1]dx+
$$

\begin{equation}
+\int\limits_{2l^+_m\leq |x|\leq L}
P_{0(\Lambda)}^{\beta}(x|x)G^0_x(X_m|Y_m)dx+\int\limits_{2l^+_m\leq |x|\leq L}
P_{0(\Lambda)}^{\beta}(x|x)G_x'(X_m|Y_m)dx\}.
\end{equation}
For $|x|\geq 2l^+_m$ we have the bound

$$
G^0_{x,e}(X_m|Y_m)\leq \frac{(2l^+_m)^3}{|x|^3}2^{4|g`|m}.
$$
Here we used the inequalities 

$$
|(\frac{x^*_j}{x^*})^{n_{1,j}^+}(\frac{x_j}{x})^{n_{1,j}^-}
(\frac{y^*_j}{x^*})^{n_{2,j}^-}(\frac{y_j}{x})^{n_{2,j}^+}|\leq
 \frac{(2l^+_m)^3}{|x|^3}2^{-(n_{1,j}^++n_{1,j}^-+n_{2,j}^++n_{2,j}^-+)},
\hskip 10 pt|C_n^{+(-)g}|\leq C^{-|g|}_n>0.
$$
After applying them we enlarge the sum in (2.5) to the sum over 
$\Bbb (Z^+)^{4m}$.

Since $P_{\Lambda}(x|x)$ tends to $(2\pi \beta)^{-1}$ the first and the second 
terms in (2.7) have limits when $L$ tends to infinity. We have only to 
calculate the third term. Let's show that

\begin{equation}
\int\limits_{r\leq |x|\leq L}G^-_x(x'|y')dx=0.
\end{equation}
For arbitrary $r,L, v=v^1+iv^2, B=B_{L}\backslash B_r$ we have 

$$
\int\limits_{B}(\frac{v}{x}-\frac{v^*}{x^*})dx=-2i[v^1\int\limits_{B}
\frac{x^2}{|x|^2}dx-v^2\int\limits_{B}\frac{x^1}{|x|^2}dx]=0,
$$

$$
\int\limits_{B}(\frac{v}{x^2}+\frac{v^*}{x^{*2}})dx=2\{v^1
\int\limits_{B}\frac{(x^1)^2-(x^2)^2}{|x|^4}dx+2v^2
\int\limits_{B}\frac{x^1x^2}{|x|^4}dx\}=0.
$$
All the above integrals are zero since the all the functions change signs 
when either a sign of one of the variables is changed , or a permutation is  
done.

As a result
\begin{equation}
\int\limits_{r<
|x|\leq L}G_x'(X_m|Y_m)dx=-\frac{1}{4}g^2(\sum\limits_{e}z_ee^2)N_{r,L}
\sum\limits_{j=1}^{m}e_j^2|x_j-y_j|^2.
\end{equation}

The integral in the right-hand-side of this equality diverges as $2ln L$.
For $g'=k, k\in \Bbb Z$ we obtain (2.4) in which we have to put $C^k_n=0$ for 
$n>k$, $C_n^k=\frac{k!}{(n-k)!n!}$.

From (2.7), (2.9) we derive

$$
G_{\Lambda}(X_m|Y_m)=\sum\limits_{e}z_e\{\int\limits_{|x|\leq 2l^+_m}
P_{0(\Lambda)}^{\beta}(x|x)[\prod\limits_{j=1}^{m}
\left(\frac{x^*-x_j^*}{x-x_j}\frac{x-y_j}{x^*-y^*_j}\right)^{\frac{1}{2}gee_j}-1]dx+
$$

\begin{equation}
+\int\limits_{2l^+_m\leq |x|\leq L}
P_{0(\Lambda)}^{\beta}(x|x)G^0_x(X_m|Y_m)dx
-\frac{1}{4}g^2(\sum\limits_{e}z_ee^2)N_{2l_m^+,L}
\sum\limits_{j=1}^{m}e_j^2|x_j-y_j|^2.
\end{equation}

This equality together with (2.8-9), in which $2l_m^+$ is substituted 
instead of $L$, yields for arbitrary $0 < r< \infty$

$$
G_{\Lambda}(X_m|Y_m)=
\int\limits_{|x|\leq r}
\sum\limits_{e}z_e
[\prod\limits_{j=1}^{m}
\left(\frac{x^*-x_j^*}{x-x_j}\frac{x-y_j}{x^*-y^*_j}\right)^
{\frac{1}{2}gee_j}-1]
P^{\beta}_{0(B_L)}(x|x)dx+
$$

$$
+\int\limits_{r<|x|\leq L}
\sum\limits_{e}z_e
[\prod\limits_{j=1}^{m}
\left(\frac{x^*-x_j^*}{x-x_j}\frac{x-y_j}{x^*-y^*_j}\right)^
{\frac{1}{2}gee_j}-1-\sum\limits_{j=1}^{m}G'_x(x_j|y_j)]
P^{\beta}_{0(B_L)}(x|x)dx-
$$

\begin{equation}
-\frac{1}{4}g^2(\sum\limits_{e}z_ee^2)N_{r,L}
\sum\limits_{j=1}^{m}e_j^2|x_j-y_j|^2.
\end{equation}

Using the equality

$$
\lim\limits_{L\rightarrow \infty}
P_{0(B_L)}^{\beta}(x|x)=
(2\pi \beta)^{-1}
$$
we deduce that the function $G$ is well defined

$$
G(X_m|Y_m)=\lim\limits_{L\rightarrow \infty}[G_{\Lambda}(X_m|Y_m)
+\frac{1}{4}g^2(\sum\limits_{e}z_ee^2)N_{r,L}
\sum\limits_{j=1}^{m}e_j^2|x_j-y_j|^2]=
$$
$$
=(2\pi \beta)^{-1}
\int\limits_{|x|\leq r}
\sum\limits_{e}z_e
[\prod\limits_{j=1}^{m}
\left(\frac{x^*-x_j^*}{x-x_j}\frac{x-y_j}{x^*-y^*_j}\right)^
{\frac{1}{2}gee_j}-1]dx+
$$

\begin{equation}
+(2\pi \beta)^{-1}\int\limits_{|x|>r}
\sum\limits_{e}z_e
[\prod\limits_{j=1}^{m}
\left(\frac{x^*-x_j^*}{x-x_j}\frac{x-y_j}{x^*-y^*_j}\right)^
{\frac{1}{2}gee_j}-1-\sum\limits_{j=1}^{m}G'_x(x_j|y_j)]dx.
\end{equation}

\vskip 20 pt
THEOREM 

Let $\Lambda$ be the ball $B_L$, centered at the origin, of radius L. Then for
 all the values of the activities $z_e$ and charges $e_j$ the thermodynamic 
limit for the RDMs exists and is zero if $x_j\not =y_j$. Moreover, the 
following equality is valid

$$
\rho(X_m|Y_m)=\lim\limits_{L\rightarrow
\infty}\exp\{g^2\frac{N_{r,L}}{4}
\sum\limits_{j=1}^{m}e_j^2|x_j-y_j|^2\}\rho^{B_L}(X_m|Y_m)=
$$
$$
=Z_{(e)_m}\exp\{i[U(X_m)-U(Y_m)]\}
P_0^{\beta}(X_m|Y_m)
e^{G(X_m|Y_m)},
$$
where $G$ is defined by (2.12) and
$$
P_0^{\beta}(X_m|Y_m)=(2\pi \beta)^{-m}\prod\limits_{j=1}^{m}
\exp\{-\frac{|x_j-y_j|^2}{2\beta}\}.
$$

\vskip 20 pt

Now, let the condition of neutrality holds $\sum\limits_{e}z_ee=0$, then
from (2.12) it follows that

$$
G(X_m|Y_m)=
(2\pi \beta)^{-1}\int\limits_{|x|\leq r} \sum\limits_{e}z_e
[\prod\limits_{j=1}^{m}
\left(\frac{x^*-x_j^*}{x-x_j}\frac{x-y_j}{x^*-y^*_j}\right)^
{\frac{1}{2}gee_j}-1]+
$$

$$
+(2\pi \beta)^{-1}\int\limits_{|x|>r} 
\{\sum\limits_{e}z_e
[\prod\limits_{j=1}^{m}
\left(\frac{x^*-x_j^*}{x-x_j}\frac{x-y_j}{x^*-y^*_j}\right)^
{\frac{1}{2}gee_j}-1]+
$$

\begin{equation}
+g^2(\sum\limits_{e}z_ee^2)\sum\limits_{j=1}^{m}e_j^2[|\frac{x_j-y_j}{x}|^2
-\frac{1}{2}
(\frac{(x_j-y_j)^2}{x^2}+\frac{(x_j^*-y_j^*)^2}{x^{*2}})]\}dx.
\end{equation}

\vskip 20 pt

From (2.13) it follows that $G(x,X_{m}|x,Y_{m})=G(X_m|Y_m)$. 
Repeating the above arguments the equality
$$
\lim\limits_{L\rightarrow\infty}(\pi L^2)^{-1}\int\limits_{B_L}
\exp\{i\sum\limits_{j=1}^{m}ee_j[\phi(x-x_j)-\phi(x-y_j)]\}dx=1
$$
is proved. So, we see that the following corollary is true.
\vskip 20 pt
COROLLARY. \hskip 20 pt If the condition of neutrality holds then the 
compatibility condition for the renormalized RDMs in the thermodynamic limit 
holds
$$
\lim\limits_{L\rightarrow\infty}\frac{2\beta}{z_eL^2}
\int\limits_{B_L}\rho(x,X_m|x,Y_m)dx=\rho(X_m|Y_m).
$$
\vskip 20 pt
AKNOWLEDGMENTS.

The research described in this publication was possible in part by the Award 
number UP1-309 of the U.S. Civilian Research and Development Foundation 
(CRDF) for independent states of the former Soviet Union.
\vskip 20 pt
REFERENCES

1 R.Jackiw,S-Y.Pi, Phys.Rev.D15,42,p.3550, 1990.
 
2 C.Trugenberger, The anyon fluid in the Bogoliubov approximation,
Preprint LA-UR 91-2421, Los Alamos, national laboratory.

3  Y.Kitazawa, H.Murayama, Nucl.Phys.,B338, p. 777, 1990.

4  J.Lykken, J.Sonnenschein, N.Weiss, Theory of anyonic superconductivity.
 A review. Preprint TAUP-1858-91, Int. Journ. of Mod.Phys., 1991.

5  W.Skrypnik, Ukrainian Math.Journ.,47, 12,p.1686, 1995.

6  W.Skrypnik, Mat.Fiz.,Analiz.,Geom., 4, N1/2, p.248, 1997.

7  W.Skrypnik, Ukrainian Math.Journ.,49, 5,p.691, 1997.

\end{document}